\documentstyle[12pt,aaspp4]{article}
\def\etal{{\it et al. }}
\def\asec{$^{\prime\prime }$}

\lefthead{Wilking et al.}
\righthead{Brown Dwarf Candidates}

\begin{document}

\title{Spectroscopy of Brown Dwarf Candidates in the $\rho$ Ophiuchi Molecular Core \altaffilmark{1}}

\author{Bruce A. Wilking \altaffilmark{2}}
\affil{Department of Physics and Astronomy, University of Missouri-St.  Louis\\
8001 Natural Bridge Road, St.  Louis, MO 63121\\ brucew@newton.umsl.edu}

\author{Thomas P. Greene \altaffilmark{2}}
\affil{Lockheed Martin Missiles \& Space, 3251 Hanover St. \\
Palo Alto, CA  94304-1191 \\
tgreene@meer.net}

\and

\author{Michael R. Meyer \altaffilmark{3}}
\affil{Steward Observatory, The University of Arizona, Tucson, AZ 85721 \\
mmeyer@gould.as.arizona.edu}

\altaffiltext{1}{Observations reported in this study were obtained at the 
Multiple Mirror Telescope Observatory, a facility operated jointly by the 
Smithsonian Institution and the University of Arizona.}

\altaffiltext{2}{Visiting Astronomer at the Infrared Telescope Facility,
which is operated by the University of Hawaii under contract from the 
National Aeronautics and Space Administration.}

\altaffiltext{3}{Hubble Fellow}

\begin{abstract}
\setcounter{page}{2}
We present an analysis of low resolution infrared spectra for 20
brown dwarf candidates in the core of the $\rho$ Ophiuchi
molecular cloud.  Fifteen of the sources display absorption-line
spectra characteristic of late-type stars.   By comparing the depths
of water vapor absorption bands in our candidate objects
with a grid of M dwarf standards, we derive spectral types which 
are independent of reddening.
Optical spectroscopy of one brown dwarf candidate confirms the spectral
type derived from the water bands.
Combining their spectral types with published near-infrared photometry,
effective temperatures and bolometric stellar luminosities
are derived enabling us to place our sample on the
Hertzsprung-Russell diagram.  We compare the positions of the brown dwarf
candidates in this diagram with two sets
of theoretical models in order to estimate their masses and ages.
Considering uncertainties in placing
the candidates in the H-R diagram, six objects consistently lie in 
the brown dwarf regime and another five objects lie
in the transition region between stellar
and substellar objects.  The ages inferred for the sample are consistent
with those derived for higher mass association members.
Three of the newly identified brown dwarfs display infrared
excesses at $\lambda$=2.2 $\mu$m suggesting that young brown
dwarfs can have active accretion disks.
Comparing our mass estimates of the brown dwarf candidates with those
derived from photometric data alone suggests that spectroscopy is an
essential component of investigations of the mass functions of young clusters.

\end{abstract}
\keywords{stars: pre-main-sequence, brown dwarfs -- infrared: stars --
ISM: individual ($\rho$ Ophiuchi cloud)}

\section{Introduction}

The shape of a star cluster's Initial Mass Function (IMF) at and
below the hydrogen burning limit can yield important information
about the star formation process as well as 
the relative importance of brown dwarfs in a localized region 
of the Galaxy.  It is becoming evident that in most clusters,
the IMF below 0.3 M$_{\sun}$
is approximately flat in logarithmic mass units and that the total 
cluster mass is not dominated by the lowest mass members
(see Luhman \etal 1998 and references therein).  Yet the detailed
shape of the IMF in clusters below 0.3 M$_{\sun}$ is poorly defined
(see Scalo 1998 for review).
To constrain the slope of a cluster IMF in this mass range, 
it will be necessary to
identify significant numbers of brown dwarfs.
This is most easily accomplished in young clusters since low mass objects
are most 
luminous in their youth.  As demonstrated in models by Burrows \etal (1997),
deuterium burning stabilizes the luminosity of a contracting brown dwarf
such that its luminosity at 10$^6$ years is about 3 orders of magnitude
greater than at 10$^9$ years.  
Recent visible wavelength studies of the
Orion Nebula Cluster (ONC, Hillenbrand 1997)
and the Taurus molecular cloud (Brice\~no \etal 1998) have indeed
been successful in detecting several young, substellar mass objects.
But due to their cool effective temperatures ($<$3000 K)
and the large columns of dust in regions of star formation, young brown
dwarfs are most easily observed at near-infrared wavelengths.
Models predict that in the closest star-forming regions (d$\approx$150 pc),
pre-main sequence (PMS) stars
of age 10$^6$ years at the hydrogen-burning limit 
will have apparent magnitudes of K (2.2 $\mu$m)=15.5$^m$ 
when viewed through A$_v$=50$^m$ and will be 
easily detectable in deep ground-based
near-IR surveys.

The $\rho$ Ophiuchi molecular cloud lies in the Upper Scorpius
subgroup of the Sco-Cen OB association.
At a distance of only 150 pc (de Zeeuw \etal 1997), 
it is an ideal
area to search for young
brown dwarfs.  It is the highest stellar density region
of low mass star formation among clouds within 200 pc of the Sun.
It hosts over 100 young stellar objects (YSOs) 
in its 1 pc $\times$ 2 pc centrally
condensed core (Grasdalen, Strom, \& Strom 1973; Vrba \etal 1975;
Wilking \& Lada 1983; Greene \& Young 1992;
Comer\'on \etal 1993; Strom, Kepner \& Strom 1995).
The central regions of the core, with A$_v$
estimated to be 50$^m$--100$^m$, provide an effective screen of background
stars even at infrared wavelengths.
The PMS stars in the core have an estimated age of
$<$10$^6$ years (Greene \& Meyer 1995) and theoretical
models predict that brown dwarfs of this age and distance 
will be easily detected at near-infrared 
wavelengths over a broad range of masses.
Previous infrared studies in the $\rho$ Oph core 
have used broadband photometry to
estimate masses of YSOs and 
identify substellar candidates 
(e.g., Rieke \& Rieke 1990; Comer\'on \etal 1993; Strom, Kepner \& Strom;
Williams \etal 1995; Comer\'on, Rieke \& Rieke 1998).  Comer\'on
and collaborators have gone a step farther and 
derived an IMF for this cluster which is relatively flat down to
to 0.05 M$_{\sun}$
(Comer\'on, Rieke, \& Rieke 1996).  Yet this result
depends on large completeness corrections at the lowest masses and,
in most cases, the mass
estimates have not been confirmed by spectroscopy.  
Using optical spectroscopy, Luhman \etal (1997)
have confirmed one of the candidates, 162349.8-242601,
as a brown dwarf with a
mass of 0.01--0.06 M$_{\sun}$.  

In this contribution, we use 
low resolution infrared spectroscopy
in the $\lambda$=2.2 $\mu$m window to identify substellar objects 
in the core of the $\rho$ Oph cloud.  The infrared spectral region is necessary
since brown dwarf candidates in this cloud will be obscured by
dust, and a spectral resolution of $R \sim 300$ is adequate given the very 
broad nature of molecular features in cool objects.
In Section 2, we describe our 
selection of brown dwarf candidates using published near-infrared photometry.
A description of how the infrared spectra were obtained for 20 $\rho$ Oph 
brown dwarf candidates
and 14 late-type dwarf standards is also given in Section 2.
Section 3 introduces a new technique, calibrated with a
grid of M dwarf standards, that uses water vapor absorption
bands in the infrared spectra to derive spectral types.
As a result, 15 brown dwarf candidates are assigned late-type
spectral classifications, with
ten objects classified as M6 or later.
The spectral classification 
for one PMS star is confirmed through optical spectroscopy.
Given their spectral types, the effective temperatures and
bolometric luminosities of the 
brown dwarf candidates are estimated in Section 4.  Using a
Hertzsprung-Russell diagram and two sets of pre-main-sequence tracks,
we find that 6 candidates clearly fall in the brown dwarf regime and
5 candidates lie in the stellar/substellar transition region.
Section 5 presents a discussion of the infrared excesses of the
probable brown dwarfs and compares our mass estimates with
those from infrared photometric studies.


\section{Observations \& Source Selection}

Low resolution infrared spectra were obtained for 
14 late-type dwarf spectral standards and 20 objects
in the core of the $\rho$ Ophiuchi molecular cloud.
Logs of the Ophiuchus observations are given in Table 1.
Included for each program object is
the date of observation, the total on-source 
integration time, and the reference star used to divide out telluric  
features.  In addition, one $\rho$ Oph object, GY~5, was observed
with the MMT Red Spectrograph.  All observations are 
described in more detail in the sections below.

\subsection{Sample Selection}

Most of the 20 brown dwarf candidates in the core of the $\rho$ Oph cloud
observed spectroscopically were selected
from the two-micron survey of Comer\'on \etal (1993).  Not only is the magnitude
limit of their survey comparable to the detection limit for our
spectroscopic observations (K$<$14.5$^m$), but
they concentrated on the highest extinction region of the cloud
core where A$_v$=50--100$^m$ and contamination 
by background stars is minimal.
We supplemented the Comer\'on \etal sample 
with sources from the near-infrared survey of Greene \& Young (1992) and an
unpublished near-infrared survey of $\rho$ Oph Core A by this group.
Candidates were chosen on the basis of their estimated absolute K
magnitudes with M(K)$\ge$3
\footnote{The absolute K magnitude, M(K), was estimated using the standard
formula M(K)=m(K)-A(K)-DM where m(K) is the apparent K magnitude,
A(K) is the extinction at K, and DM=5.9$^m$ which is appropriate for the
Ophiuchus cloud.  The extinction at K was estimated from A(K)=1.4 $\times E(H-K)$
(Cohen \etal 1981) assuming an intrinsic $(H-K)_0$=0.3 typical for an
M dwarf and no infrared excess.  
For comparison, a PMS star of age 3 $\times$ 10$^5$ years and M=0.10 M$_{\sun}$
is predicted to have a M(K)=4.2 (D'Antona \& Mazzitelli 1998).}.
In other words, we preferentially selected sources which were faint at
K and had values of $(H-K)\le$2.
The small $(H-K)$ colors favor PMS stars near the surface of the cloud
and/or with small
infrared excesses which minimizes the effects of spectral veiling.

A $K \ vs. \ $(H-K) diagram for the 20 brown dwarf candidates is shown in
Fig. 1 relative to the zero-age main sequence
and a 3 $\times$ 10$^5$ year old isochrone.
The median values of apparent K magnitude, $(H-K)$ color, and absolute K
magnitude for our 20 objects are K=12.4$^m$, $(H-K)$=1.3$^m$, and M(K)=4.5$^m$.
Among the 14 objects in the Comer\'on \etal survey with K$<$14$^m$
and M(K)$>$4$^m$,
we obtained infrared spectra for ten.
Our selection criteria appear to have been successful at picking PMS stars in
the cloud, and only 3 objects are suspected background stars (Sec. 3.3).
The heavily reddened source GY~31 was observed unintentionally, as it fell
into the slit during an attempt to observe GY~30.

\subsection{IRTF Grism Observations}
All infrared spectra were obtained between 1996 June and 1998 January 
with the 3.0 m NASA IRTF on Mauna Kea.  Observations were made using
the facility infrared camera, NSFCAM,
in the K band with the HKL grism, a 0.3 arcsec pixel$^{-1}$
scale, and a 0.6\asec\ slit (see Shure \etal 1994 and Rayner \etal 1998
for a description of the instrument and grism). 
The spectra cover the 2.0--2.5 $\mu$m
band and have a resolution R=$\lambda/\Delta\lambda\sim$300.
Flat field exposures of the dome interior illuminated with incandescent lamps
were taken at the beginning of each night.
A typical set of observations consisted of the star
observed at two positions along the slit, separated by 10 arcsec.  
Typical exposure times for the brown dwarf candidates
were 120--180 seconds per slit position
and limited by variations
in the atmospheric OH emission.  Total on-source integration times 
were typically 2 minutes for the standard star observations (but extended
for 9 minutes for the M9 stars) and 18 minutes for the brown dwarf
candidates (see Table 1).
Before extraction, the spectral images in each set were sky-subtracted
using their companion image,
and divided by a flat-field.  Wavelength calibration was established
using OH emission lines.  After extraction, telluric features
in the spectra were removed by dividing by
the spectrum of a dwarf star near spectral type A0 and  
observed typically within 30--40 minutes and
within 0.1 airmasses of the program objects.  The airmass difference
between the telluric standard and program star never exceeded 0.25.
No attempt was made to restore the true continuum shape of
the program objects; to recover the true shape
one could multiply each spectrum by a blackbody
corresponding to the effective temperature of the telluric standard.
Because of the steep response function of the grism, it was not
possible to extrapolate over the Br $\gamma$ absorption line in the
telluric standard and hence we were not sensitive to
Br $\gamma$ emission in the program stars.
The telluric standard star used for each brown dwarf candidate
is listed, along with their spectral types,
in the last column of Table 1.

\subsection{MMT Red Spectrograph}

A low resolution optical spectrum was obtained for GY~5 under 
non--ideal conditions on 1998 February 13,
at the Multiple Mirror Telescope on Mt. Hopkins.  We used the Red 
Channel spectrograph with the 270 line mm$^{-1}$ grating and a 2\asec\
wide slit resulting in a resolution of 23.5 $\AA$ per spectral
element over the range 6500--9500 $\AA$.
The detector was a 1200 $\times$ 800 
pixel Loral CCD which was binned $\times 2$ in the spatial dimension.  
Data were taken in beam--switching mode, nodding the source along the slit
with integration times of 600 seconds per exposure.  
Image processing was accomplished by subtracting beam--switched
pairs from each other to remove bias, dark current
and sky emission.  The subtracted frames were flat--fielded and corrected for
grating efficiency using a set of calibration frames obtained with a 
featureless quartz lamp.
Wavelength calibration was achieved with a comparison spectrum taken
with a HeNeAr lamp. The individual spectra were then extracted 
and the local sky level was re--calculated 
using apertures adjacent to that containing the source 
spectrum.  The four spectra were median filtered, 
and the resultant spectrum was smoothed with a gaussian of 
half-width $\sigma = 11.5 \AA$ for comparison with the 
library of spectral standards from Kirkpatrick, Henry, \& McCarthy (1991).

\section{Results}

Infrared spectra of 13 standard stars, ranging in spectral
type from K5V to M9V, are shown in Fig. 2 with the most 
prominent atomic and molecular
absorption features labeled.  Spectral classifications were adopted from
Kirkpatrick, Henry, \& McCarthy (1991) or Henry, Kirkpatrick, \&
Simons (1994).  Also shown is a spectrum of the
star BRI 0021-0214 which has tentatively been classified as $>$M9.5V
(Kirkpatrick, Beichman, \& Skrutskie 1997).  
Apparent in the dwarf spectra are narrow absorption lines
by Na~I (blended doublet at 2.21 $\mu$m), Ca~I (blended triplet 
at 2.26 $\mu$m), 
and the CO $\Delta\nu$=2 bands
(2.29-2.42 $\mu$m).  Not labeled are
broad absorptions by H$_2$O vapor at the beginning and end of the bandpass.
Sections of the spectra have been removed that span the wavelengths
of the Br$\gamma$ absorption line (2.166 $\mu$m) in the telluric standard, and
where division of atmospheric features was poor ($\sim$2.03 $\mu$m).
ASCII data files of the 14 standard star spectra and a table of
the observational details are electronically available 
upon request.

\subsection{Spectral Classifications Using Water Vapor}

At this low spectral resolution, there are no absorption lines 
present in our spectra that
can be reliably used to derive spectral types.  However,
the depths of the broad water vapor absorption bands are extremely sensitive to
spectral type for M dwarfs.  We have derived
a quantitative index that measures these depths
which is independent of reddening.  This index is analogous to the Q index
derived from UBV photometry which is used to determine spectral types 
for visible stars (e.g., Johnson \& Morgan 1953).
Using the average values of relative flux density calculated in
three narrow bands, F1 (2.07--2.13 $\mu$m), F2 (2.267--2.285 $\mu$m), 
and F3 (2.40--2.50 $\mu$m),
we have defined a reddening-independent water index as
\begin{equation} 
Q = (F1/F2)(F3/F2)^{1.22} 
\end{equation}
assuming a reddening law from 1.6--3.5 $\mu$m of A$_\lambda$=$\lambda^{-1.47}$
(Rieke \& Lebofsky 1985, Cohen \etal 1981).
A plot of the H$_2$O index Q vs. spectral type is shown in Fig. 3 for the M
dwarf standards shown in Fig. 2.
The error bars in the Q values are calculated from the statistical errors
in the average flux values of F1, F2, and F3.  The plot shows that
Q appears to be linear from spectral type M0 to M8 but may start to flatten
by M9.  
A weighted linear least squares fit to the M standard star data
(excluding the K5V star Gl~775 and $>$M9.5V star BRI 0021-0214) 
data yields the following relation:
\begin{equation}
 MV~subclass = (-18.35\pm1.72) \times Q + (17.00\pm0.45)
\end{equation}
with a correlation coefficient of r=0.98.
We expect that this relation cannot be used for spectral classification
of objects with very strong CO bands, such as those found in late-type
giants, as the higher vibrational first overtone bands become evident
in the 2.4-2.5 $\mu$m region.
The application of this relation to spectral classifications
of the brown dwarf candidate spectra is discussed in the following section.

\subsection{Spectral Classifications of Brown Dwarf Candidates}

Photospheric absorption lines characteristic of cool, late-type stars were
observed in 15 of the 20 spectra of $\rho$ Oph objects.  
However, placing these objects on the H--R diagram
requires a two--step process; i) deriving spectral types for individual
objects from comparison with spectral standards; and ii) adopting a
temperature scale appropriate for the sample in question.
The spectral 
classification of these objects is described below in Sec. 3.2.1.
The effects on the derived spectral types
of spectral veiling by circumstellar dust and of surface gravities 
less than those of dwarf stars 
are discussed in Sec. 3.2.2 and Sec. 3.2.3, respectively.
We defer discussion of the effects of
surface gravity on our adopted temperature scale to Sec. 4.
Of the remaining 5 objects lacking evidence for late-type photospheres,
GY~30 displayed a featureless spectrum and WL~18 has
Br$\gamma$ emission in an otherwise featureless spectrum (see also
Greene \& Lada 1996).  The S/N in the spectrum
for GY~218 (K=14.4) was insufficient to attempt any classification.
Finally upon further analysis,
two objects (CRBR~1 and 2316.6-2131) showed Br$\gamma$ absorption characteristic
of stars of spectral type earlier than G0.  These objects are discussed in
Sec. 3.3.3.

\subsubsection{Analysis of Water Vapor Features}

Fifteen of the $\rho$ Oph objects have spectra displaying absorption
by CO, Na~I, Ca~I, and H$_2$O as expected from late--type photospheres.
Their composite CO equivalent widths 
(sum of $\nu$=0-2 and 2-4) range from $>$6.5 $\AA$ (GY~59) to $>$25 $\AA$
(GY~107) with a median value of 12.5 $\AA$.  With the possible 
exception of GY~107,
these equivalent widths are consistent with those observed in
M dwarf stars.  However, they are inconsistent with CO equivalent
widths expected in M giants with strong water vapor absorption bands
(e.g., Fig. 7 of Greene \& Lada 1996).
CO band absorption in the 2.4--2.5 $\mu$m region is present in
GY~107, and may cause the Q index and the photospheric temperature
to be underestimated.   For most sources in our sample, which appear
to be late-type dwarfs, the CO absorption from 2.4--2.5
$\mu$m has a negligible effect on the water-vapor index.
Thus we conclude that most of the 15 objects in our sample are 
candidate very low mass stars and that our dwarf star calibration
of the Q index introduced above is appropriate for estimating their
spectral types.

We assume that the observed absorption features arise from some
combination of a YSO photosphere plus possible contribution
from continuum dust emission and not from a disk photosphere.
While optically--thick continuum dust emission is commonly
observed toward actively accreting T Tauri stars (Strom \etal 1989;
Beckwith \etal 1990),
optically--thick gas emission is thought to dominate near--IR emission
only for objects with very high disk accretion rates such as the
FU Ori class of eruptive variables (Kenyon \& Hartmann 1996).
Comparison of the CO equivalent widths for our candidate brown dwarfs
with the sample of FU Ori stars studied by Greene \& Lada (1996)
at low resolution suggests that
none of our objects belong to this rare class of objects.
However, water vapor absorption can also arise in circumstellar disks
surrounding less extreme PMS stars under certain conditions
according to the models of Calvet \etal (1992).
Features due to water vapor can contribute as much as 10\% of
the observed flux for accretion rates 10$^{-8}$ M$_{\odot}$ yr$^{-1}$
only if the circumstellar disk extends all the way into the stellar
surface.  At an accretion rate of 10$^{-7}$ M$_{\odot}$ yr$^{-1}$, water
vapor becomes significant only for disks that extend to within 2 R$_*$.
Note that these calculations assume a central stellar temperature
of 4000 K.  A cooler central star will have a colder disk,
making it less likely that inner disk will reach temperatures
where gas opacity is dominant ($>$ 1000 K).
Whether it is because disk
accretion rates are typically $< 10^{-7} M_{\odot}$ yr$^{-1}$
(Hartmann \etal 1998),
or inner disks have inner holes $> 2 R_*$ (Meyer \etal 1997),
it appears that the near-IR spectra of
most YSOs are dominated by star+continuum dust
emission 
rather than disk photospheres (Casali \& Eiroa 1996; Greene \& Lada 1997).

We have computed the reddening-independent 
water vapor index Q for each object in our sample using Eqn. 1.
Using the relation between the water vapor index and MV subclass
(Eqn. 2), spectral types were computed for each brown
dwarf candidate.
We considered whether or not our water vapor index could have
been influenced by our telluric correction procedure.
An analysis of the derived Q values as a function of airmass
difference between program object and telluric standard
shows no correlation.
Considering the uncertainties in the Q values and in the coefficients
of eqn. 2, the spectral types are thought to be accurate to $\pm$1.5 subclasses.
Q values and spectral types are presented in Table 2.  The derived
spectral types vary from M2.5 to M8.5.  Ten of the objects have spectral
types of M6 or later.  Spectra of these objects are presented in
Figures 4a and 4b.  
For comparison, we display spectra of M dwarf standards,
artificially reddened using a $\lambda^{-1.47}$ law, that provide
a good match to the shape of the continuae of the brown dwarf candidates.

The optical spectrum of GY~5, shown in Fig. 5, 
gives us an opportunity to check the 
viability of the spectral types determined for the reddened
brown dwarf candidates using the water vapor index.  
We compared our optical spectrum of GY~5 with dwarf and giant star 
standards 
of Kirkpatrick \etal (1991).  Based on the relative strength of surface 
gravity sensitive 
features (e.g., Na~I at $\lambda$ 8183 $\AA$), 
GY~5 appears to be intermediate between luminosity class III and V as 
expected for a PMS star sitting above the main sequence.  
Similar behavior is 
reported by Luhman \etal (1997) for another brown dwarf candidate identified 
in the Ophiuchus dark cloud and observed with the same instrument.  
However, based upon
the behavior of molecular features such as TiO and VO, 
GY~5 appears to be closer to the dwarf sequence.   As shown in Fig. 5,
a comparison of our spectrum with the dwarf standards of Kirkpatrick
\etal (1991) suggests a spectral type of M6$\pm$1.0 for GY~5.  This 
agrees well with the spectral type derived 
above using the Q index (M7$\pm$1.5).  We note the absence H$\alpha$
in emission with the EW(H$\alpha$)$<$ 5 \AA.

\subsubsection{Effects of Veiling}

Although the computed H$_2$O index Q is insensitive to reddening by 
interstellar extinction, other effects may cause variations in Q 
between stellar standards and PMS stars of identical spectral type. 
One major concern alluded to above 
is wavelength-dependent infrared excess due to 
thermal emission from warm dust grains in the circumstellar environments of 
the PMS stars.  
A geometrically--thin, optically--thick
circumstellar disk is expected to exhibit
an IR spectral shape characterized by $F_{\lambda}
\propto \lambda^{-7/3}$, whether heated via viscous accretion
or passive reprocessing (Lynden-Bell \& Pringle 1974; 
Adams, Lada, \& Shu 1987).
We adopted this function to simulate the 
spectrum of a circumstellar disk at each observed wavelength in our 
spectral bandpass.  We divided the resulting spectrum by a T=9000 K 
Planck function in order to process it comparably to the observed spectra 
which have been divided by the spectrum of a dwarf star (usually
near A0V) for telluric correction.

What is the magnitude of the effect of thermal emission by dust on spectral
types derived using the water vapor index Q?
K band excess emission can be characterized by 
$r_{k} = F_{Kex}/F_{K*}$, where $F_{Kex}$ is the broadband
flux from excess (circumstellar) 
emission and $F_{K*}$ is stellar flux at 2.2 $\mu$m.  Meyer \etal (1997)
found that classical T Tauri stars have a median value 
of $r_{k} \simeq 0.6$, while weak-emission T Tauri stars have a
median value of $r_{k} \simeq 0.0$. 
In order to simulate this 
range of excesses, we scaled a normalized disk spectrum to equal 0.2 
and 0.6 times the flux
of several standard star spectra at $\lambda = 2.20 \mu$m. 
These scaled model disk spectra were then added to those of 
the standard stars, resulting in spectra which are representative of 
PMS stars whose K band emissions arise from both stellar photospheres and 
disks.  We then recomputed the Q indices of these resultant spectra 
and found that Q increased (implying earlier spectral types) over 
that of the same standards without this excess emission added.  Both 
the M6.5V (GJ~1111) and M9V (LHS~2924) standards appeared to have 
Q-derived spectral types 1 subclass earlier when $r_{k} = 0.2$ disk 
spectra were added, and spectral types 3 
subclasses earlier when $r_{k} = 0.6$ disk spectra were added.

We have estimated the minimum likely K band excess, $r_{k}$,
for each brown dwarf candidate from published near-infrared photometry
and include these results in Table 2.  These estimates are 
minima because our calculations assume that the candidates have no excesses 
in the $J$ or $H$ bands.  The K excess is estimated
from $r_{k} = 10^{0.4 \times E(H-K)_0} -1$, where $E(H-K)_0 = (H-K) - 
0.065A_v - (H-K)_0$ (e.g., Meyer \etal).  The value of $r_{k}$ is
less than 0.3 for all sources except for CRBR~15, GY~11, GY~64, and GY~202, 
which have values similar to those of classical T Tauri stars.  
Thus the Q-derived 
spectral types of most brown dwarf candidates in the sample 
should be unaffected by emission from circumstellar dust, 
but CRBR~15, GY~11, GY~64, and GY~202 may have actual spectral 
types which are 2--3 subclasses later than those presented in Table 2. 

\subsubsection{Surface Gravity Effects}

We have compared the spectra of our brown dwarf candidates to a sequence
of dwarf standards rather than giants.  Yet
surface gravities for PMS stars are known to be intermediate
between those of dwarfs and giants (e.g., Schiavon, Batalha, \& Barbury 1995).
Spectroscopic studies indicate that the surface gravities of PMS stars are
most similar to those of main sequence dwarfs (Basri \& Batalha 1990;
Greene \& Lada 1997).  This is consistent with our analysis of the 
composite CO equivalent widths in our sample and the GY~5 spectrum.
If the brown dwarf candidates have surface gravities of subgiants,
how would this effect our derived spectral types?  
Absorption in the 1.9 $\mu$m water vapor band is generally weaker 
in late-type giants than in
dwarf stars of the same spectral type
(e.g., Aaronson, Frogel, \& Persson 1978, Kleinmann \& Hall 1986).
Only for the latest
spectral types ($>$M6) is the water band similar in depth for
dwarfs and giants.  Hence, we may have systematically 
classified stars 1--2 subclasses earlier than appropriate for stars $<$M6 
by assuming dwarf rather than subgiant surface gravities.
Classifications of stars M6 or later appear to be nearly
independent of surface gravity.  
Higher resolution data,
obtainable with the next generation of 8--10 m class
telescopes, are required in order to perform a detailed analysis
of the surface gravities of very low mass PMS objects.
Note that there is still an
open question of surface gravity effects on our translation of spectral
types to effective temperatures.  We defer this discussion to Sec. 4.

\subsection{Notes on Individual Sources}


\subsubsection{GY~11}
The infrared spectrum of GY~11 (Fig. 4b) shows absorption bands of CO
and H$_2$O, modified by the presence of dust emission.  Its infrared
excess is confirmed through weak $\lambda$=10 $\mu$m emission 
(Rieke \& Rieke 1990) and recent ISOCAM observations 
(Comer\'on \etal 1998).
In addition, 
emission lines due to H$_2$ are also present: the 1$\rightarrow$0 S(1)
at 2.122 $\mu$m and the 1$\rightarrow$0 Q-branch lines at 2.407, 2.413,
and 2.424 $\mu$m.  Emission from the 1$\rightarrow$0 S(1) transition has
been previously reported by Williams \etal (1995).
Such emission line spectra are characteristic of YSOs
with molecular outflows (Reipurth \& Aspin 1997;
Greene \& Lada 1996) and could be 
evidence that the low-luminosity PMS star GY~11 has an associated outflow.
A second possibility is that the H$_2$ emission is due to a chance coincidence
between GY~11 and an H$_2$ emission knot from the VLA~1623 outflow.  As seen in
deep H$_2$ images, 
there is a ``stream"
of H$_2$ emission between VLA~1623 and H$_2$ knot H4 that intersects
the position of GY~11 (see Fig. 7 of Dent, Matthews, \& Walters 1995).

\subsubsection{Possible Background Stars CRBR~1 \& 2316.6-2131}

While not shown here, both
CRBR~1 and 2316.6-2131 display spectra with Br $\gamma$ absorption
and no CO bands.  This suggests thay have
spectral types of G0 or earlier (e.g., Ali \etal 1995; Wallace \& Hinkle 1997). 
Adopting intrinsic colors appropriate for this inferred range of spectral types,
neither object exhibits an infrared excess at K; their color excess ratios, 
$E(J-H)/E(H-K)$, are consistent with the value of $1.57 \pm 0.03$
derived for the 
Ophiuchus cloud by Kenyon, Lada, \& Barsony (1998).
>From their observed $(H-K)$ color,
the visual extinctions 
toward CRBR~1 and 2316.6-2131 are estimated to be A$_v\sim$22$^m$ and 
$\sim$25$^m$, respectively.  The observed slopes of the continuae 
in their
infrared spectra are consistent with the spectrum of an early-type star
reddened by these amounts.  Given the early spectral types and lack
of infrared excess, it is possible that CRBR~1 and 2316.6-2131
are background stars viewed through the dark cloud.  Indeed, the
visual extinctions we estimate are consistent with the total cloud extinction
of A$_v$=36$\pm$18$^m$ estimated from the C$^{18}$O column densities
along the lines of sight to each source (Wilking \& Lada 1983).  

\subsubsection{Possible Background Giant GY~107}

The infrared spectrum of GY~107 shows the signatures of a late-type
photosphere including strong CO band absorptions.
Near-infrared photometry indicates it does not have an infrared excess. 
The visual extinction estimated to this source of A$_v$=13$^m$
is consistent with
the total cloud extinction of 
15$\pm$8$^m$ estimated from the
C$^{18}$O column density along this line of sight.
Therefore it is possible that GY~107 is a background star viewed
through the dark cloud.  The strengths of the CO bands are consistent
with a K5--M0III classification and not the M3V classification
derived from the Q index (e.g., Fig. 4 in Greene \& Lada 1996).
A giant classification would imply
a distance greater than 1 kpc, consistent with being a background 
star.  

\section{Masses and Ages of the Brown Dwarf Candidates}

To place the brown dwarf candidates on a Hertzsprung-Russell (H--R) diagram
and estimate their masses and ages requires us first to convert their
spectral types to effective temperatures and to determine
their bolometric luminosities. 
The adopted temperature scale for M dwarfs is based on the modified
blackbody fits of Jones \etal (1996) and is derived in Appendix A.
Effective temperatures computed for each brown dwarf candidate are
given in Table 2 and have typical 
uncertainties in the Log of $\pm$0.035 dex, dominated by the uncertainty
in computing each value of Q.
However, systematic effects could affect the derived effective temperatures.
As alluded to in Sec. 3.2.2, effective temperatures for
CRBR~15, GY~11, GY~64, and GY~202 could be overestimated by 300-500K
due to the presence of veiling.
The assumption of dwarf, rather than subgiant, surface gravities could
also systematically effect our results.
Because late--type giant star temperatures are
300-500 K higher than the adopted dwarf temperature scale
(compare Table 5 with the giant temperature scale in Perrin \etal 1998),
we may be underestimating the effective temperature by $\sim$150-250 K
for stars later than M6. For stars classified as M2--M6, the
temperatures quoted here 
should not be systematically off since the higher effective temperature
scale for subgiants is compensated for by the later
spectral type needed to produce the same water vapor absorption (Sec. 3.2.3).
With these caveats in mind, we adopt the temperatures listed in Table
2 for our brown dwarf candidates.

Given their spectral types, bolometric luminosities were
estimated for each brown dwarf candidate by i) dereddening
their J magnitudes using the corresponding intrinsic
$(J-H)$ colors for M dwarfs; ii) applying bolometric corrections
to the J magnitudes appropriate for each spectral type; and iii) 
converting absolute magnitudes to bolometric luminosities using  
the standard relation Log(L$_{bol}$/L$_{\sun}$) = 1.89 - 0.4 $\times$ 
M$_{bol}(*)$.  Infrared photometry for the
brown dwarf candidates is presented in Table 2 and is from the study of 
Barsony \etal (1997) in the CIT photometric system.
Intrinsic colors and bolometric corrections as a function of MV subclass 
were derived using the studies of Leggett (1992), Tinney, Mould,
\& Reid (1993) and Leggett \etal (1996); 
the derivation of these values is described in Appendix A.
For two sources without J photometry, bolometric luminosities were
estimated in the same manner from the H and K photometry.
Visual extinctions could be overestimated for 
objects with excess emission at H resulting in overestimates for
Log(L$_{bol}$/L$_{\sun}$) by about 0.25 dex (0.1 dex) assuming $r_{h}$=0.6
($r_{h}$=0.2). 
Values for Log(L$_{bol}$/L$_{\sun}$)
are presented in Table 2 and have formal uncertainties of
$\pm$0.16 dex.

H--R diagrams for the 15 brown dwarf candidates are  
shown in Figure 6 for two sets of theoretical tracks.
Fig 6a. shows their placement relative to the evolutionary tracks and isochrones
of D'Antona \& Mazzitelli (1998) from 0.02 and 0.3 M$_{\sun}$.
The models use opacities from Alexander
\& Ferguson (1994) and the Full Spectrum of Turbulence convection model
of Canuto \& Mazzitelli (1991) and assume mass fractions of helium, metals, 
and deuterium of
Y=0.28, Z=0.02, and X$_D$=2$\times$10$^{-5}$.
These models are similar to those 
described in D'Antona \& 
Mazzitelli (1997), but with an improved treatment of deuterium burning.
Fig. 6b shows the candidates on a H--R diagram alongside 
the theoretical tracks of Burrows \etal (1997) for masses between 0.01 and 0.24
M$_{\sun}$.
These models are constructed specifically for substellar objects and
consider a variety of opacity sources, particularly water
vapor, and assume mass fractions of helium, metals, and deuterium of
Y=0.25, Z=0.02, and X$_D$=2$\times$10$^{-5}$.

The masses and ages derived for our brown dwarf candidates from both models
are presented in Table 3.
Regardless of choice of tracks, one sees that the majority 
of objects do appear to have
ages of 10$^6$ years or less, with a median age of 3--4 $\times$ 10$^5$
years.  These ages agree with those derived for more massive cluster
members by 
Greene \& Meyer based on the D'Antona \& Mazzitelli (1994) tracks
\footnote{The D'Antona \& Mazzitelli (1998) tracks used in our analysis
are identical to their 1994 models for M$>$0.2 M$_{\sun}$.}.
This agreement gives one confidence in the adopted models
but one must bear in mind that while the relative ages may be robust,
absolute ages are model dependent and 
difficult to verify by independent means
(see for example Fig. 5 in Luhman \& Rieke 1998).
The apparent older age for GY~11 is most likely
a result of its veiled photosphere which leads us to overestimate its 
effective temperature.
{\it Six of the brown dwarf candidates, CRBR~14, GY~10, GY~11, GY~64, 
GY~202, and GY~310, consistently fall
below the hydrogen burning limit of 0.08 M$_{\sun}$, taking into account
the uncertainties in their effective temperatures.}
To be consistent in age with the other candidates, GY~11 must have a mass in 
the range $<$0.02 M$_{\sun}$.  In addition, 
the brown dwarf candidates
CRBR~15, GY~5, GY~37, GY~59, and GY~84 lie in the transition region
between stellar and substellar objects and we cannot rule out
that some are brown dwarfs. 
Again, masses derived from PMS models require independent confirmation by 
studies of PMS binary stars (Ghez \etal 1995) or other
techniques (Bonnell \etal 1998).
The range of masses derived from both sets
of tracks considered here 
gives some indication of the uncertainties in the mass estimates
(see Table 3).

\section{Discussion}

In the previous section, we demonstrated that six of our
candidate objects appear to be brown dwarfs according to
the PMS tracks considered.  An additional five stars in
our sample lie at the stellar--substellar boundary at an
age of $< 10^6$ yrs.  What can we say about the nature of
these very low mass objects?  By studying these and other
recently discovered young brown dwarfs in more detail, we
hope to learn whether very low mass stars are assembled in
a similar fashion to higher mass stars.  We conclude with a
discussion of how purely photometric techniques compare with our
mass estimates based on spectroscopy.  

\subsection{Infrared Excesses for Substellar Objects}

Do young brown dwarfs 
exhibit infrared excess emission
characteristic of active accretion disks?
If PMS stars of masses 0.5--1.0
M$_{\sun}$ evolve from 0.1 M$_{\sun}$ protostars with massive accretion
disks, perhaps the lowest mass association members are those that form
without significant disk accretion.
In our sample, CRBR~15, GY~11, GY~64, and GY~202 have obvious 2.2 $\mu$m
excesses (lower limits are given in Table 2 as r$_k$).  
This group includes three of our six probable brown dwarfs.
These 2.2 $\mu$m excesses are confirmed by combining near-infrared
data and our effective temperature estimates with ISOCAM observations
in the thermal IR (Comer\'on \etal 1998). 
In addition, an excess is suggested for
CRBR~14 at wavelengths beyond 5 $\mu$m, indicating the
presence of an inner hole in the circumstellar dust distribution.
Finally we note that the $\rho$ Oph
brown dwarf studied by Luhman \etal (1997) exhibits strong
H$\alpha$ emission as well as thermal IR excess emission, suggesting
the presence of an accretion disk.  In summary, indications are 
that brown dwarfs
in the Ophiuchus dark cloud can indeed possess active accretion
disks analogous to those found around higher mass PMS stars.

Using excess emission at $\lambda$=2.2 $\mu$m to identify circumstellar 
disks among members of the ONC, Hillenbrand et al.
(1998) find a lower disk frequency for the lowest mass cluster
members (M$<$0.2 M$_{\sun}$); a disk frequency of only 20--50\%
is estimated for substellar objects compared to 55--90\% 
for the cluster as a whole.
Hillenbrand et al. forward several possible
explanations for this lower frequency including smaller disk masses,
lower disk accretion rates, and/or a higher incidence of disk stripping
for the lowest mass cluster members.  
By identifying a statistically significant sample of brown dwarfs
in the core of the Rho Oph cloud, this latter hypothesis can be tested in a
cluster of approximately the same age as the ONC.  Stellar encounters
and the stripping of circumstellar disks would be less prevalent
in the lower stellar density environment of the Rho Oph core
whose stellar density is a factor of $\sim$50 times lower
than the ONC density of $\sim$5 $\times$ 10$^4$ pc$^{-3}$.


\subsection{Comparisons of Spectroscopic and Photometric Techniques}

Even though spectroscopy affords the most reliable way to estimate effective
temperatures of stars and young brown
dwarfs, it has limitations.  
First, while optical spectroscopy provides
the best studied sets of lines for spectral
classification, the visual extinction in a star-forming cloud severely limits
the number of young brown dwarfs that can be observed at $R \sim 1000$
in the R and I bands.  Among the 15 brown dwarf candidates
in this study, only six (with extinctions less than 13$^m$) 
appear in an I band image of the cloud
(I$<$19$^m$, Wilking \etal 1998) and are accessible targets for
moderate resolution spectroscopy on a 4-m class telescope.
Second, even infrared spectroscopy is currently limited
to low-luminosity objects near the surface of the cloud.  In our study,
the probable brown dwarfs range in extinction from 5--14$^m$ while
the total extinction in the core is estimated to be 50--100$^m$.
Brown dwarfs throughout
most of the cloud volume are below the detection limit even for low
resolution infrared spectroscopy with a 3--4 m class telescope.

Ultimately, any study of the mass functions of young clusters 
down to substellar
masses will have to combine sensitive photometric surveys with 
spectroscopic data.  The main disadvantage of analyzing photometric 
data alone is that mass and age cannot be uniquely determined.  
Typically one either assumes an input mass distribution and derives
an age from fits to PMS evolutionary tracks (e.g., Lada \& Lada 1995) 
or one assumes an age distribution and derives the mass function 
(e.g., Comer\'on, Rieke, \& Rieke  1996).   
The addition of spectroscopic information permits one to construct 
an H--R diagram for some fraction of the photometric sample in order
to estimate the age distribution of the cluster, crucial for adopting 
the appropriate mass--luminosity relationship (e.g., Meyer 1996). 
Given this age distribution, the derived distribution 
of stellar masses should reflect the cluster mass function 
for a large statistical sample. 

How do determinations of mass made with infrared spectra 
compare with those determined from infrared photometric data?
For example, the color-magnitude diagram in Fig. 1 based upon
the D'Antona \& Mazzitelli models would predict
that for an age of 3 $\times$ 10$^5$ years, 15 of the 20 brown dwarf 
candidates (75\%)
have masses $\le$0.1 M$_{\sun}$.   Using the infrared spectra and the 
same input models, we
have determined that 11 of 14 (79\%) candidates have M$\le$0.1 M$_{\sun}$,
with three of the brown dwarf candidates identified as possible
background 
stars.  One of the important advantages of infrared spectroscopy
is the ability to reduce contamination of the sample by field stars.

A recent photometric study by Comer\'on \etal (1998) used model fits of
the SEDs of Ophiuchus brown dwarf candidates,
extended to mid-infrared wavelengths by ISOCAM data at $\lambda$=
3.6 $\mu$m, 4.5 $\mu$m, and 6.0 $\mu$m, to estimate effective
temperatures.  They utilized the Burrows \etal
(1997) models to estimate mass assuming an age of either 10$^6$ or
3 $\times$ 10$^6$ years.  The distribution of sources in our 
H--R diagram (Fig. 6) suggests that ages ranging from 10$^5$ 
to 10$^6$ years would be more appropriate and 
would tend to give lower mass estimates than those quoted by 
Comer\'on \etal  Seven objects are in common
between this study and Comer\'on \etal, including five of the six probable brown
dwarfs identified in our study.
In general, the temperatures derived by their 
SED models are systematically higher
by several hundred K compared to those suggested from the infrared spectra,
but usually within the stated uncertainty of our technique.  Notable
exceptions include GY~10 and CRBR~14; the photometric technique derives
an effective temperature for GY~10 which is 700 K hotter than indicated
from the near-infrared spectrum and 350 K hotter for CRBR~14.
Masses derived by the Comer\'on \etal study are typically a factor of two 
greater than we derive from the same PMS tracks (Burrows \etal models; 
see col. 4-5, Table 3) with larger differences evident for CRBR~14,
CRBR~15, and GY~10.   In the case of CRBR~15, the higher mass derived
by Comer\'on \etal (0.23 M$_{\sun}$ vs. $<$0.09 M$_{\sun}$ from this study)
arises from using a higher visual extinction than we estimate from the
(J-H) color and spectral type.  Their model appears to overestimate the 
luminosity of the star while underestimating the contribution to the 
bolometric luminosity from infrared excess emission.   
In summary, our mass estimates usually agree
within a factor of two of those derived by SED fitting, although they are 
systematically lower as expected given that the objects appear to be cooler
and slightly younger than assumed by Comer\'on \etal

\section{Summary}

We have obtained low resolution infrared spectra for 20 brown
dwarf candidates which lie toward the core of the $\rho$ Ophiuchi
dark cloud and for 14 late-type spectral standards.  Analysis of these spectra,
in combination with published near-infrared photometry, has enabled
us to derive effective temperatures and bolometric luminosities
for the brown dwarf candidates.  Utilizing the PMS evolutionary
models of Burrows \etal (1997) and D'Antona \& Mazzitelli (1998),
we estimate masses and ages for the sample, as well as
consider the distribution of circumstellar material around these
low mass young stars.
Our analysis has
yielded the following results:

$\bullet$ We have developed a novel technique that uses the infrared 
water vapor bands to derive spectral
types for M dwarfs, independent of reddening, to within $\pm$1.5 subclasses.
An optical spectrum obtained for one brown dwarf candidate, GY~5,
confirms the spectral classification derived
from the water bands.

$\bullet$ Fifteen of the brown dwarf candidates display photospheric
absorption features characteristic of late-type stars and all but one
appear to be low mass members of the $\rho$ Ophiuchi cloud.
The derived spectral types range from
M2.5--M8.5.  The presence of veiling in four of the objects leads
us to derive spectral types which are too early by 2--3 subclasses.
The brown dwarf candidate GY~107 is identified as a
possible background giant.  

$\bullet$ An H--R diagram of the brown dwarf candidates indicate
that six are substellar
in nature for both of the evolutionary models considered here.
Another five candidates lie in the transition region between stellar
and substellar objects.  
The median age for the sample is 3--4 $\times$ 10$^5$
years, consistent with the ages determined for more massive PMS objects
in the cloud.

$\bullet$ Three of the probable brown dwarfs display infrared excesses
at $\lambda$=2.2 $\mu$m and four have excesses at mid-infrared wavelengths.  
By analogy with classical T Tauri stars in the Taurus dark cloud,
the near--infrared colors of these objects suggest that young
brown dwarfs in this cloud can have active accretion disks.


Future studies of the emergent mass distribution in $\rho$ Oph
will require combining spectroscopic studies such as this, with
deep infrared imaging photometry in order to constrain the mass
function below the hydrogen burning limit.  Follow--up observations
aimed at characterizing the distribution of circumstellar material
surrounding these very young brown dwarfs
will be extremely useful in constraining models of very low
mass star formation.

\acknowledgements 
We would like to thank Jim Liebert, Kevin Luhman, 
and George Rieke for important discussions that guided the
interpretation of these data.  We especially thank John Carpenter 
who assisted in
formulating the candidate list and Joan Najita who suggested using
water vapor bands for spectral classification.
We are grateful to Kevin Luhman and Lynne Hillenbrand for providing 
electronic versions of the 
Kirkpatrick et al. standards and assistance with the spectral 
classification of GY~5.  David Sing and Ale Milone 
provided expert assistance in obtaining the spectrum of GY~5 
during the last hour of the last night of the last regularly scheduled
run of the old MMT.  Mary Barsony generously provided
infrared photometry in advance of publication.
We thank the IRTF and telescope operators Bill Golisch
and Charlie Kaminski for performing service observations for
two standard stars.
B.W. gratefully acknowledges the support of RUI Grant NSF AST-9417210
to the University of Missouri-St. Louis.  Support for MRM was provided 
by NASA through Hubble Fellowship grant \# HF-01098.01-97A awarded by
the Space Telescope Institute, which is operated by the Association of
Universities for Research in Astronomy, Inc., for NASA under contract
NAS 5-26555.

\appendix
\section{Adopted Temperature Scale, Intrinsic Colors, and 
Bolometric Corrections for M Dwarfs}

In order to use infrared spectra and broadband near-infrared photometry
to place our brown dwarf candidates in an H-R diagram, we need to adopt a
temperature scale, intrinsic $(J-H)_0$ and $(H-K)_0$ colors,
and bolometric corrections at J (1.25$\mu$m) and K (2.2$\mu$m)
for each spectral type.
No single modern compilation of these values exists for M dwarfs, so
we developed our own based on recent data presented in the literature.
For consistency, we have used the M dwarf spectral classifications
of Kirkpatrick, Henry, \& McCarthy (1991) and Henry, Kirkpatrick, 
\& Simons (1994).
To minimize the effects of metallicity, only stars classified as
disk objects based on kinematics were considered (Leggett 1992).
Our temperature scale for M dwarfs was established using the modified
blackbody fits in Table 3 of Jones \etal (1996).  Their stated random error
in the determination of temperature scales is $\Delta$T$_{eff}$ = 150 K.
We chose their temperatures over those derived by
other studies as they included three dwarfs of very late spectral type
(M8, M8.5 and M9) and did not rely on fitting of synthetic spectra
which currently
have trouble reproducing spectral features in very late M dwarfs.
Intrinsic $(J-H)$ and $(H-K)$ colors
were taken from Leggett (1992) and Leggett \etal (1996) and are
in the CIT photometric system.  Bolometric corrections at K 
were taken from Leggett \etal (1996) and Tinney, Mould, \& Ried (1993) and
are also in the CIT system.  Bolometric
corrections at J were derived using the (J-K) colors and $M_K$. 
Table 4 summarizes the temperatures, intrinsic colors, and
bolometric corrections compiled for individual stars.

To derive a temperature scale, and continuous intrinsic $(J-H)$
and $(H-K)$ colors and bolometric corrections as a function of M 
spectral class, uniform-weighted fits were
made to the data in Table 4 over the spectral range M2V--M9V.
A linear least
squares fit to the Jones \etal temperatures yields the following 
temperature scale:
\begin{equation}
T_{eff} = (-166\pm8) \times (MV~subclass) + (3758\pm52)  K.
\end{equation}
A linear fit to the $(J-H)$ and $(H-K)$ color vs. MV~subclass
yields the following relations: 
\begin{equation}
(J-H)_0 = (0.0187\pm0.0047) \times (MV~subclass) + (0.503\pm0.026)
\end{equation}
\begin{equation}
(H-K)_0 = (0.0437\pm0.0023) \times (MV~subclass) + (0.096\pm0.012) .
\end{equation}
A linear fit to the K bolometric corrections, BC$_K$, results
in the following relation:
\begin{equation}
BC_K = (0.0895\pm0.0065) \times (MV~subclass) + (2.43\pm0.04)  .
\end{equation}
A second order fit was needed for the J bolometric corrections, BC$_J$,
as the peak flux of the M2-M9 dwarfs moves completely through
the J band.  The fit yielded:
\begin{equation}
BC_J = (-0.0105\pm0.0034) \times (MV~subclass)^2 + (0.148\pm0.040)
 \times (MV~subclass) + (1.53\pm0.10) . 
\end{equation}

\noindent While the $(J-H)_0$ dependence with spectral 
type appears double-valued over a 
broader range of spectral types (e.g., Bessell \& Brett 1988), 
it is best represented by a linear fit
over our range of MV~subclasses.
of wavelength, it is best fit by 
The adopted values for effective temperature, $(J-H)_0$, $(H-K)_0$, BC$_J$, and
BC$_K$ are given in Table 5 from M2V to M9V in increments of 0.5
subclasses.  

\pagebreak

\begin{center}
{\bf Figure Captions}
\end{center}
\medskip
%
%
\figcaption{%
A color-magnitude diagram of the Ophiuchus brown dwarf candidates.  
All photometry
is taken from the study of Barsony \etal (1997).  Isochrones for stars of
masses ranging from 0.04--2.5 M$_{\sun}$ were derived from the models of
D'Antona \& Mazzitelli (1994, M$>$0.2 M$_{\sun}$) and D'Antona \&
Mazzitelli (1998, M$\le$0.2 M$_{\sun}$) in the CIT photometric
system using intrinsic colors and bolometric corrections for
dwarf stars derived in Appendix A.
Isochrones are shown for 3 $\times$ 10$^5$ years 
and 5 $\times$ 10$^8$ years.
Reddening lines from the 3 $\times$ 10$^5$ year isochrone are
shown for selected masses by dashed lines and were calculated using the
extinction law derived by Cohen \etal (1981).  For comparison, a reddening
vector for an A$_v$=10$^m$ is also shown.
}
%
\figcaption{%
Infrared spectra of M dwarf standards at R=300.  Spectral types were adopted
from Kirkpatrick, Henry, \& McCarthy (1991) or Henry, Kirkpatrick,
\& Simons (1994).  Fig. 2a shows the 2.0-2.5
$\mu$m spectra for dwarfs from K5 to M4.  Fig. 2b shows spectra for dwarfs
of spectral type M4.5 to $>$M9.5.  The wavelengths of absorption
lines of Na, Ca, Mg, and CO are marked at the bottom of each plot.  
The wavelength
ranges averaged to compute the water vapor index Q are labeled as F1, F2,
and F3.
}
%
%
\figcaption{%
A plot of the water vapor index, $Q$, vs. MV subclass for our M dwarf
standards.  Spectral types were adopted from Kirkpatrick, Henry, \& McCarthy
(1991).  Error bars reflect one-sigma uncertainties in $Q$
from the statistical uncertainties in calculating F1, F2, and F3.  
The solid line
is the best fit to the data (Eqn. 2).
}

%
%
\figcaption{%
Infrared spectra of selected brown dwarf candidates in the $\rho$ Oph
cloud at R=300.  The water vapor index indicates these objects have spectral
types $\ge$M6.  Fig. 4a shows the 2.0-2.5
$\mu$m spectra for candidates that most closely resemble the spectrum of 
the lightly reddened
M7 dwarf shown at the bottom of the stack.  The M dwarf spectrum was
produced by artificially reddening our spectrum of Gl 644c by A$_v$=5$^m$
assuming A$_{\lambda}\propto\lambda^{-1.47}$.  
Fig. 4b shows spectra for candidates most closely resembling the spectrum of
the more heavily reddened M6 dwarf shown at the bottom of the stack.
The M dwarf spectrum was produced by artificially reddening our spectrum of
Gl 406 by A$_v$=10$^m$.
The wavelengths of absorption
lines of Na, Ca, Mg, and CO are marked at the bottom of each plot.  The wavelength
ranges averaged to compute the water vapor index Q are labeled as F1, F2,
and F3.
}
%
%
\figcaption{%
An optical spectrum for the brown dwarf candidate GY~5 relative to
dwarf spectral standards of Kirkpatrick \etal (1991).  The
spectrum of this PMS object appears intermediate to that of a
M5.5 and M6 dwarf. Labeled on
the plot are photospheric atomic absorptions due to K~I, Na~I, and 
Ca~II and molecular absorptions from TiO and VO.  Features
labeled as O$_2$ are telluric in nature.  For a more complete
identification of features, see Kirkpatrick \etal (1991), Fig. 5c.
}
%
\figcaption{%
Hertzsprung-Russell diagrams for the $\rho$ Oph brown dwarf candidates.
Fig. 6a shows the candidates relative to the theoretical tracks
of D'Antona \& Mazzitelli (1998).  Fig. 6b shows the candidates relative to the
theoretical tracks of Burrows \etal (1997).  Isochrones from
10$^5$ years to 10$^8$ years are shown by solid lines and evolutionary tracks
from 0.02 M$_{\sun}$ to 0.30 M$_{\sun}$ are shown by dashed lines.
The bold dashed line marks the evolutionary track for a star at the 
hydrogen-burning limit.
The typical error bar
for a candidate is shown in the lower left of each plot and is
$\pm$0.035 dex in Log(T$_{eff}$) and $\pm$0.16 dex in 
Log(L$_{bol}$/L$_{\sun}$).
}

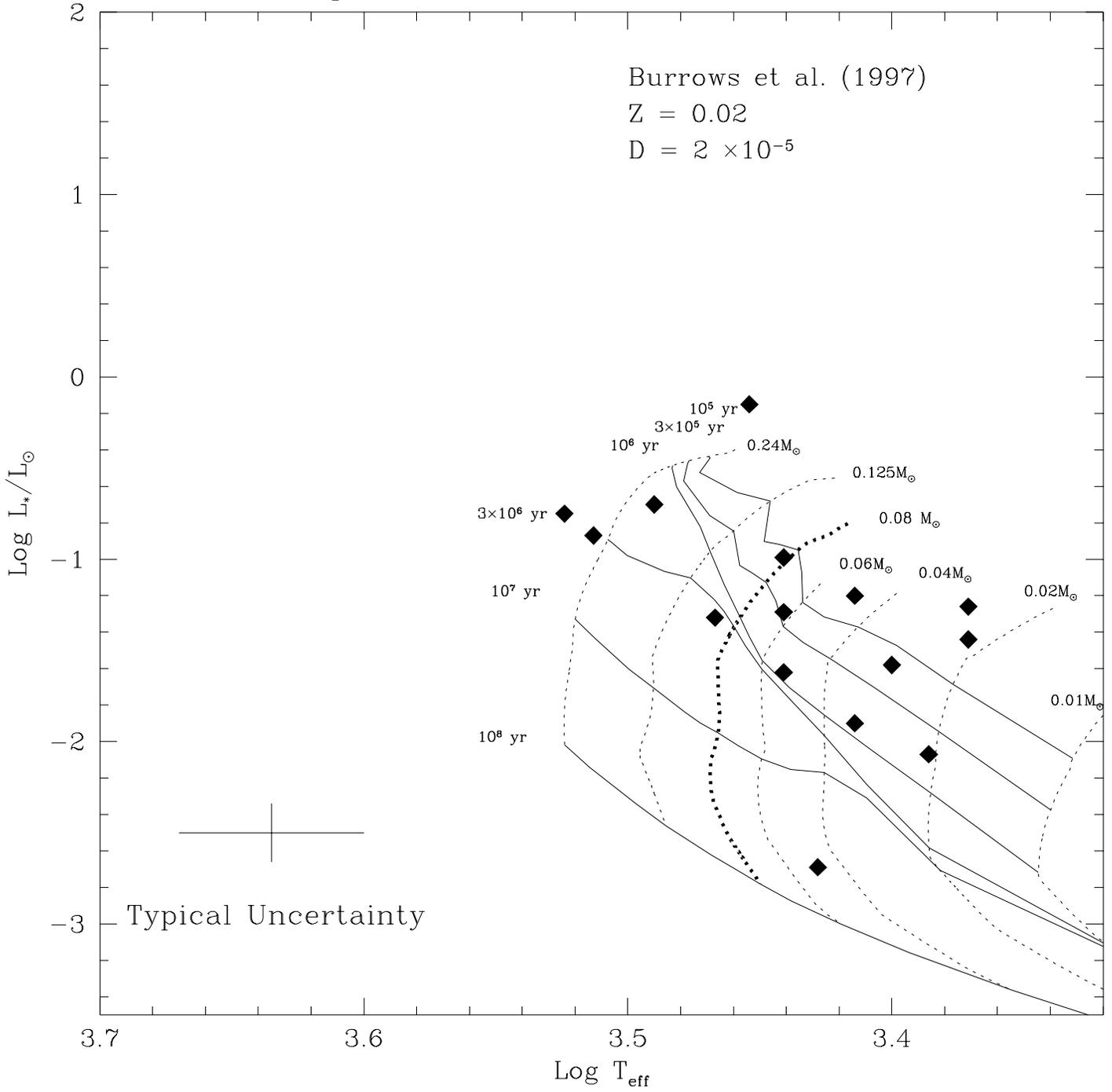
\end{document}